\documentclass[sigconf]{acmart}

\usepackage{multirow} % Table formatting
\usepackage{svg} % Handling .svg figures
\usepackage[inline]{enumitem}
\usepackage{mleftright}

\AtBeginDocument{%
  }

\copyrightyear{2024}
\acmYear{2024}
\setcopyright{rightsretained}
\acmDOI{XXXXXXX.XXXXXXX}

\acmConference[CONSEQUENCES@RecSys ’24]{In CONSEQUENCES Workshop at RecSys ’24}{October 14th, 2024}{Bari, Italy}
\acmISBN{978-1-4503-XXXX-X/18/06}

\begin{document}

\newcommand{\todo}[1]{\textcolor{red}{\textbf{[TODO: #1]}}}
\newcommand{\parahead}[1]{\noindent \textbf{#1.}}

\bibliographystyle{ACM-Reference-Format}

\title{The Relevance of Item-Co-Exposure For Exposure Bias Mitigation}
\titlenote{Based on a full paper published in the ACM TORS\citep{krause2024mitigating}}
    
\author{Thorsten Krause}
\email{thorsten.krause@dfki.de}
\affiliation{%
  % \institution{German Research Center for Artificial Intelligence}
  \institution{German Research Center for AI}
  % \institution{DFKI}
  \city{Osnabrück}
  % \state{Lower Saxony}
  \country{Germany}
}

\author{Alina Deriyeva}
\email{alina.deriyeva@dfki.de}
\affiliation{%
  % \institution{German Research Center for Artificial Intelligence}
  \institution{German Research Center for AI}
  % \institution{DFKI}
  \city{Osnabrück}
  % \state{Lower Saxony}
  \country{Germany}
}

\author{Jan H. Beinke}
\email{jan.beinke@dfki.de}
\affiliation{%
  % \institution{German Research Center for Artificial Intelligence}
  \institution{German Research Center for AI}
  % \institution{DFKI}
  \city{Osnabrück}
  % \state{Lower Saxony}
  \country{Germany}
}

\author{Gerrit Bartels}
\email{gerrit.bartels@uni-osnabrueck.de}
\affiliation{%
  \institution{University of Osnabrück}
  \city{Osnabrück}
  % \state{Lower Saxony}
  \country{Germany}
}

\author{Oliver Thomas}
\email{oliver.thomas@dfki.de}
\affiliation{%
  % \institution{German Research Center for Artificial Intelligence}
  \institution{German Research Center for AI}
  % \institution{DFKI}
  % \city{Osnabrück}
  % \state{Lower Saxony}
  % \country{Germany}
  % \institution{University of Osnabrück}
  \city{Osnabrück}
  % \state{Lower Saxony}
  \country{Germany}
}

\renewcommand{\shortauthors}{Krause et al.}

\begin{abstract}
Through exposing items to users, implicit feedback recommender systems influence the logged interactions, and, ultimately, their own recommendations. 
This effect is called exposure bias and it can lead to issues such as filter bubbles and echo chambers. 
Previous research employed the multinomial logit model (MNL) with exposure information to reduce exposure bias on synthetic data. 

This extended abstract summarizes our previous study in which we investigated whether (i) these findings hold for human-generated choices, (ii) other discrete choice models mitigate bias better, and (iii) an item's estimated relevance can depend on the relevances of the other items that were presented with it. 
% To rule out unobserved confounders, w
We collected a data set of biased and unbiased choices in a controlled online user study and measured the effects of overexposure and competition. 

We found that (i) the discrete choice models effectively mitigated exposure bias on human-generated choice data, (ii) there were no significant differences in robustness among the different discrete choice models, and (iii) only multivariate discrete choice models were robust to competition between items.
We conclude that discrete choice models mitigate exposure bias effectively because they consider item-co-exposure. 
Moreover, exposing items alongside more or less popular items can bias future recommendations significantly and item exposure must be tracked for overcoming exposure bias.
We consider our work vital for understanding what exposure bias it, how it forms, and how it can be mitigated.
\end{abstract}

%    <concept>
%        <concept_id>10002951.10003317.10003347.10003350</concept_id>
%        <concept_desc>Information systems~Recommender systems</concept_desc>
%        <concept_significance>500</concept_significance>
%        </concept>
%  </ccs2012>

% \ccsdesc[500]{Information systems~Recommender systems}

% \keywords{Exposure Bias, Discrete Choice Modeling, Exponomial Choice Model}


\maketitle

\section{Introduction}
% \todo{Mindestens eines der Modelle im Text formalisieren}
Exposure bias arises when a previous recommendation policy influenced the logged users' choices \citep{liu2020general}. 
It is a property of implicit feedback data sets that affects future models during training \citep{chen2023bias}.
Existing approaches debias learning from binary datasets that only contain positive user-item interactions but do not differentiate between rejects or non-observations \citep{zhang2021causal, lee2022bilateral, saito2020unbiased}.
An alternative approach is changing how data are collected. 
If the core problem is the unavailability of exposure information, then we could simply collect this information alongside the user-item interactions.
We would then only need to find a way for processing it.

This extended abstract summarizes our study on using discrete choice models for considering choice alternatives and mitigating exposure bias \citep{krause2024mitigating}.
Previous research proposed replacing the loss-functions of existing models with discrete choice models \citep{train2009discrete} that can consider observed alternatives as well as the (sometimes fuzzy) transitivity through rationality that holds over multiple choices \citep{ccapan2022dirichlet, krause2022beyond}.
Assuming that random utility maximization \citep{train2009discrete} accurately captures users' behavior, we can condition their choices on the choice sets, dropping the effect of exposure. 
However, the only studies on this approach relied on synthetic data that fulfilled the standard MNL's assumptions \citep{ccapan2022dirichlet, krause2022beyond}.
Empirical evidence suggests that the MNL can be inaccurate for modeling human choices \citep{blattberg1989price, benson2016relevance}.
Hence, we defined our first research question:
\textbf{RQ1.:} \textit{Does the MNL reduce exposure bias on human choice data compared with traditional recommendation approaches?}
Further, we asked:
\textbf{RQ2.:} \textit{Can other discrete choice models reduce exposure bias more than the MNL?}
% Wir testen auf Daten echter Menschen

% Andere Modelle könnten genauer sein, unklar 

% Wir testen mehrere Modelle

Moreover, existing approaches only consider items' \emph{exposure frequencies} as a source of exposure bias. 
When recommendations include multiple items, an item's choice probability can depend on the alternatives due to reference-point effects \citep{abeler2011reference} or mututally exclusive choices.
Hence, which items are exposed together, i.e., \emph{competition}, should also be considered a form of exposure bias.
Because binary data lack this information, no model that learns from binary data, including existing debiasing models, could counter this effect. We asked:
\textbf{RQ3.:} \textit{Can the composition of exposed items induce bias, and which models can mitigate it compared with exposure bias through non-uniform exposure frequencies?}

To measure the effects of exposure on discrete choice-based models and baselines, we collected a partially biased dataset in a controlled online user study, and analyzed how different exposure policies affect estimated item ranks.
The discrete choice-based models mitigated nearly all exposure bias while out-performing the baselines in accuracy. 
Moreover, we found that only the discrete choice-based models mitigated exposure bias from competition. 
Our implementation is publicly available on GitHub\footnote{\url{https://github.com/krauthorDFKI/DiscreteChoiceForBiasMitigation}}.
\section{Experimental Setup}
We aimed to measure how shifting the exposure distribution w.r.t.
% Our goal was to measure the effect of shifting the exposure distribution on recommendations regarding 
(i) \emph{exposure frequencies} and (ii) \emph{competition} affects recommendations.
We elaborate on our procedure here as we deem it potentially useful to others and are unaware of any similar existing methods.

\begin{figure}[tb]
    \centering
    \includegraphics[width=0.85\columnwidth]{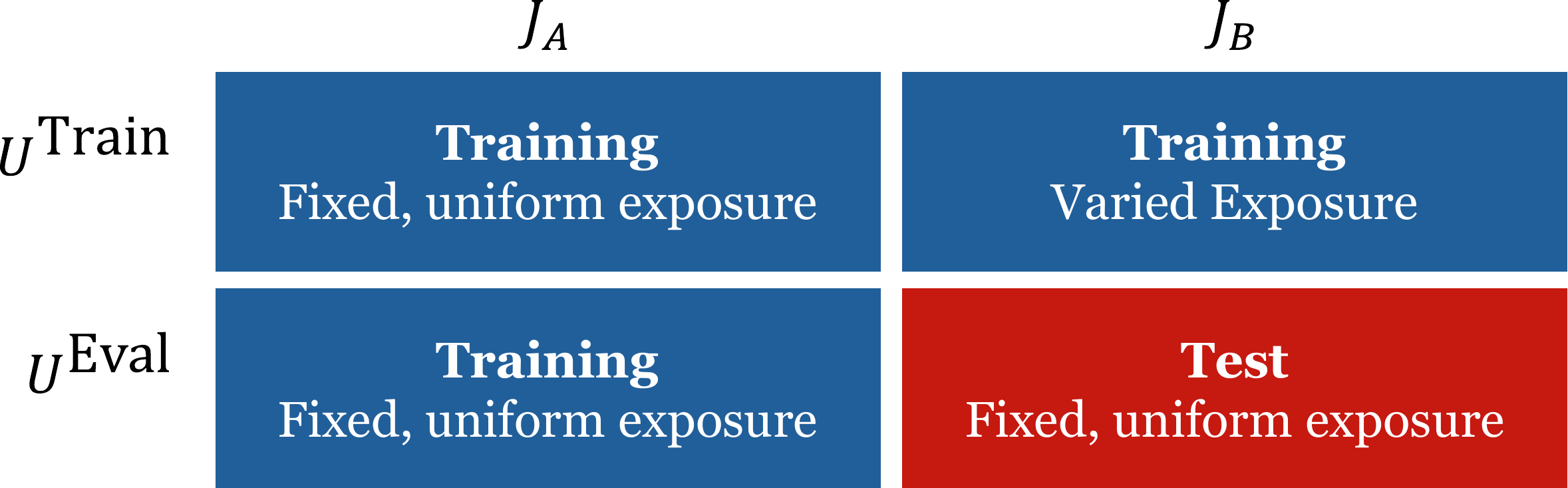}
    \vspace{-0.5\baselineskip}
    \caption{
    Dataset structure.
    }
    \label{Fig. Synthetic dataset structure.}
    \vspace{-\baselineskip}
\end{figure}

Figure \ref{Fig. Synthetic dataset structure.} illustrates the dataset structure. First, we collected partially biased choices in an online survey. 
The item set consisted of 100 educational courses split into two 50-course subsets $J_A$ and $J_B$.
We drew a five-item subset $J_\mathrm{Bias}$ from $J_B$. 
Each participant made 40 choices from four-item choice sets of which
50\% were uniformly sampled from $J_A$, 25\% were uniformly sampled from $J_B$, and 25\% were sampled from $J_B$ but more likely to contain an item from $J_\mathrm{Bias}$.

We then compiled dataset pairs from the choices so that each pair only differed in the exposure distribution on $J_B$. 
The distributions were either (i) overexposing $J_\mathrm{Bias}$ versus uniformly exposing all items or (ii) exposing the items from $J_\mathrm{Bias}$ with popular versus unpopular competitors. 
Crucially, we split the user set into two subsets $U^\mathrm{Train}$ and $U^\mathrm{Eval}$ and did not train on any choices of the users from $U^\mathrm{Eval}$ on $J_B$. 
This enabled distilling the effect of some users' exposure on other users' recommendations while avoiding information leakages. 
For measuring (i) overexposure, we corrected for some items appearing more popular on some datasets by chance. % We refer to the original contribution \citep{krause2024mitigating} for more detail.

\section{Results and Discussion}

\subsection{Exposure frequency}
Figure \ref{Fig. Bias overexposure.} compares the bias from (i) overexposure between the models, i.e., how much higher the models ranked items due to overexposure, on average over 500 simulation repetitions.

Clearly, all negative sampling-based approaches suffered strongly from exposure bias except for CPR \citep{wan2022cross}. 
Overexposing an item caused them to overestimate its rank by up to 17, which equals 34\% of $J_B$. 
Note that we arbitrarily selected the rate of overexposure and cannot conclude on the real-world effect size.
CPR was the least biased model. 
However, Figure \ref{Fig. Performance overexposure.} shows that it also performed close to the Random model in terms of nDCG. 
In the main contribution, we discussed that some changes to negative sampling, that our dataset's special structure required, could have corrupted CPR's performance, which relies on a complex and unique sampling strategy.
Among the debiasing baselines, only MACR \citep{wei2021model} and BISER \citep{lee2022bilateral} noticeably reduced bias, by about 33\% of the other models.

Importantly, the discrete choice-based models exhibited almost no bias. 
The MNL, GEV and BL underestimated the overexposed items' popularity slightly.
Hence, including information about observed alternatives can remove exposure bias from overexposure almost entirely.
Even the univariate BL model was able to deal with this source of exposure bias.
From our observations, we answered \textbf{RQ1.} as follows: 
\textit{Yes. The MNL reduced exposure bias on human choice data more than all baselines [...].}

\subsection{Competition}
For exposure bias from (ii) competition, the picture was similar but not identical.
Negative sampling-based approaches were again more biased than the discrete choice-based approaches.
This time, the bias attributed for 8 to 10 item ranks, which equals up to 20\% of set $J_B$.
Again, this number does not necessarily apply for real-world use cases.
However, no debiasing baselines noticebly reduced bias, except for CPR, which we excluded from the analysis as discussed above. 
Previous research had only focused on exposure bias from (i) exposure frequencies, apparently neglecting (ii) competition. 

Strikingly, this time, only the multivariate models exhibited almost no bias. 
BL was significantly biased, though less than the other univariate models.
Hence, we concluded, the multivariate property matters for adressing exposure bias from competition and confirmed \textbf{RQ3.}: \textit{Yes, we were able to induce bias by varying the competitiveness of choice sets. [...] Only models that consider the choice set mitigated it without sacrificing predictive accuracy. }

\subsection{Differences between discrete choice models}
Within the selection of discrete choice models, we could not spot any significant differences in bias robustness or performance.
This suggests that either the precise specification of the discrete choice model does not matter or that the dataset was too small for measuring significant effects.
Because we lack a theoretical foundation for why one discrete choice model could be more or less biased than another, we leave this matter for future research.
This led us to answering \textbf{RQ2.} with: 
\textit{We did not find conclusive evidence that other discrete-choice models could reduce bias further. Future studies 
should compare the models in more detail for a definite conclusion.}
\section{Conclusion}
We summarized our previous study on how incorporating discrete choice models into implicit feedback recommender systems can mitigate exposure bias \citep{krause2024mitigating}.
We found that we can effectively reduce exposure bias by logging observed but not chosen items and considering them during training with discrete choice models. Moreover, we showed that the composition of choice sets is also a source of exposure bias that previous approaches neglected.

Our findings imply that logging policies should also track choice alternatives. Moreover, the real-world effect of exposure bias could be enormous. In our experiment, we overexposed items by a factor of $3.2$. On real-world datasets, exposure discrepancies can be much higher, for example on the \emph{last.fm} dataset \citep{abdollahpouri2020multi}. 
Hence, adjusting logging policies and implementing discrete choice models could fundamentally change recommendations.
However, how logging and processing choice alternatives affects computational complexity compared to negative sampling remains unclear. 
Also, our findings have yet to be reproduced on real-world data. Some datasets that include item-co-exposure are the \emph{finn.no} dataset \citep{eide2021finn} and this year's RecSys challenge dataset from \emph{Ekstra Bladet}\footnote{\hyperlink{https://recsys.eb.dk/dataset/}{https://recsys.eb.dk/dataset/}}.
Shortly after ours, another study considered inter-item effects on choice from a causal perspective \cite{ruiz2024ranking}.
Given the increasing attention on this matter, we believe that our study shines important light on why choice alternatives must be considered, and on how framing the interaction with recommendations as a choice setting achieves just that.

%\begin{acks}
%To Robert, for the bagels and explaining CMYK and color spaces.
%\end{acks}

\bibliography{sample_base}

\appendix

\section{Measured bias}

\begin{figure}[H]
    \centering
    \includegraphics[width=\columnwidth]{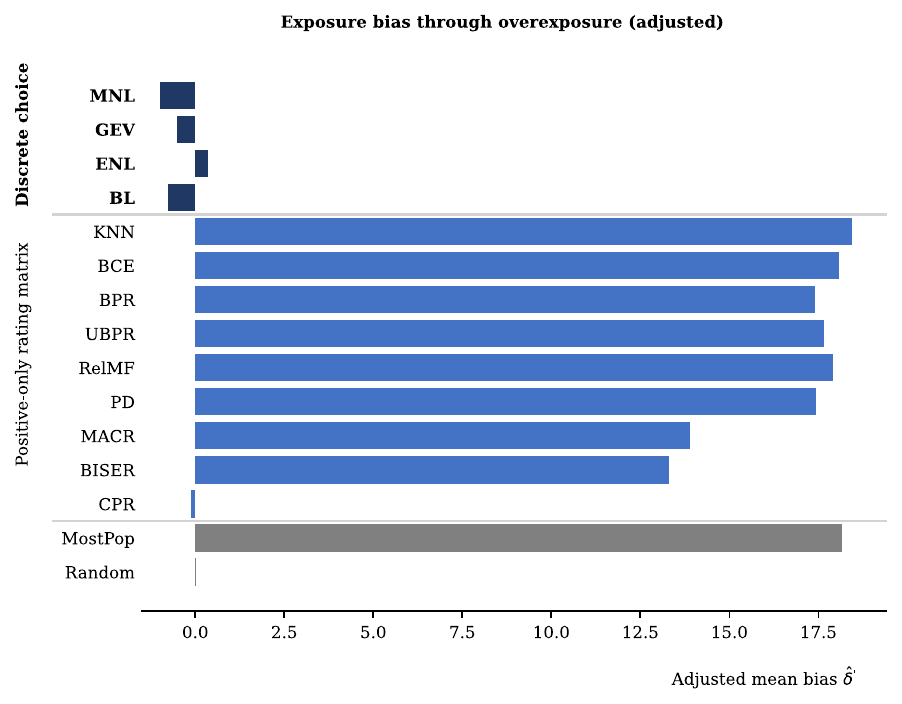}
    \vspace{-1.5\baselineskip}
    \caption{
    Bias from overexposure from \citep{krause2024mitigating}.
    }
    \label{Fig. Bias overexposure.}
    \vspace{-\baselineskip}
\end{figure}

\begin{figure}[H]
    \centering
    \includegraphics[width=\columnwidth]{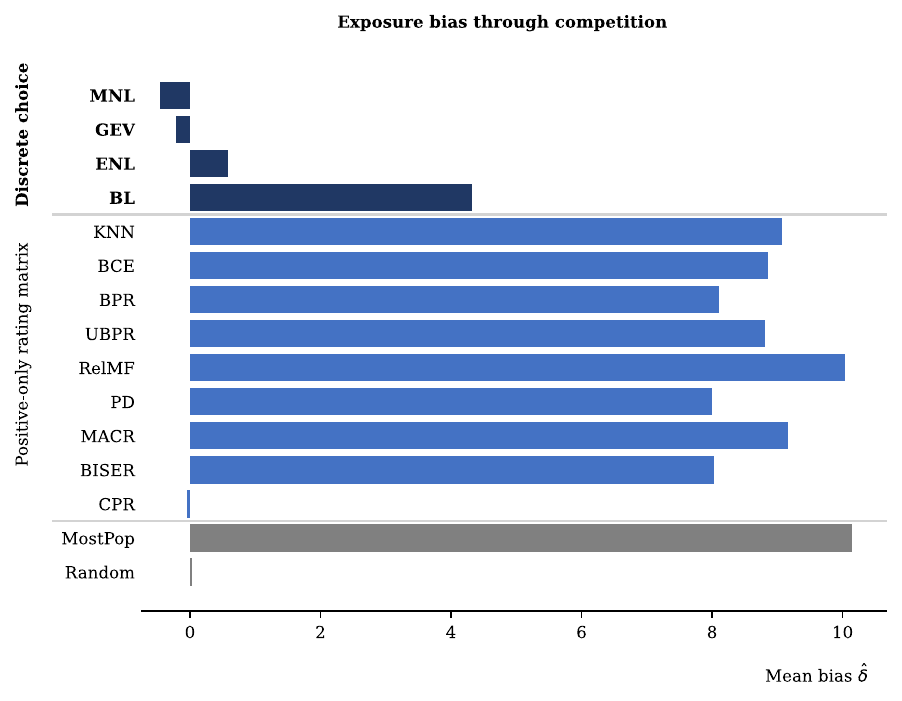}
    \vspace{-1.5\baselineskip}
    \caption{
    Bias from competition from \citep{krause2024mitigating}.
    }
    \label{Fig. Bias competition.}
    \vspace{-\baselineskip}
\end{figure}

\vspace{\fill}
\pagebreak[0]
\vspace{-\fill}

\section{Measured performance}

\begin{figure}[H]
    \centering
    \includegraphics[width=\columnwidth]{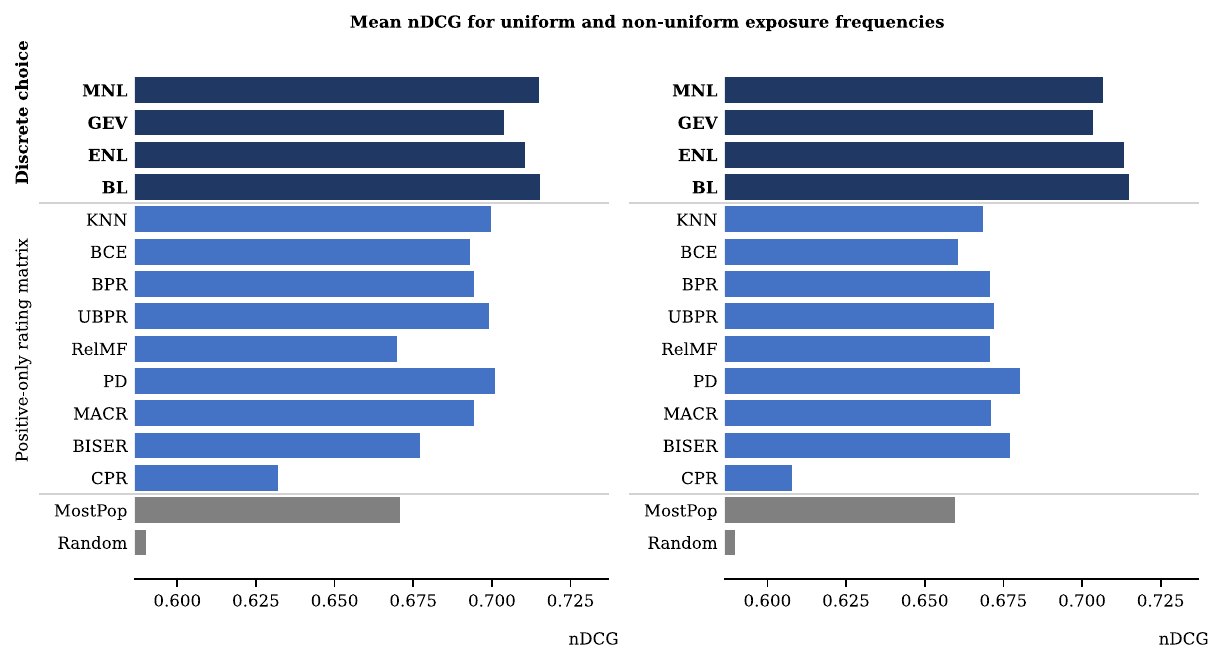}
    \vspace{-1.5\baselineskip}
    \caption{
    Measured nDCG when training on data with uniform exposure frequencies and overexposure from \citep{krause2024mitigating}.
    }
    \label{Fig. Performance overexposure.}
    \vspace{-\baselineskip}
\end{figure}

\begin{figure}[H]
    \centering
    \includegraphics[width=\columnwidth]{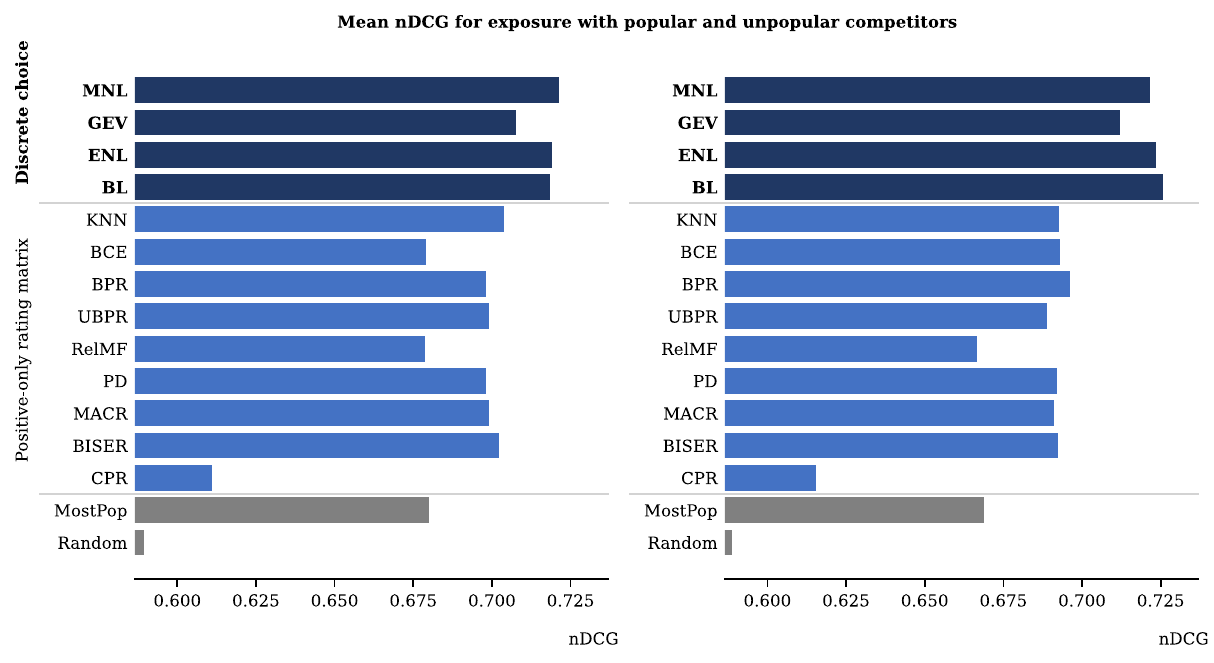}
    \vspace{-1.5\baselineskip}
    \caption{
    Measured nDCG when training on data with popular and unpopular competition from \citep{krause2024mitigating}.
    }
    \label{Fig. Performance competition.}
    \vspace{-\baselineskip}
\end{figure}

\end{document}